\documentclass{elsart}

\usepackage{amssymb}

\begin{document}

\begin{frontmatter}



\title{Modeling stylized facts for financial time series}


\author{M.I. Krivoruchenko$%
^{*,\dagger )}$, E. Alessio$^{\dagger )}$, V. Frappietro$^{\dagger )}$, L.J. Streckert$^{\dagger )}$}
\address{
$^{*)}$Institute for Theoretical and Experimental Physics, B.
Cheremushkinskaya 25\\
117259 Moscow, Russia\\
$^{\dagger )}$Metronome-Ricerca sui Mercati Finanziari, C. so Vittorio
Emanuele 84\\
10121 Torino, Italy
}

\begin{abstract}
Multivariate probability density functions of returns are constructed in order to model 
the empirical behavior of returns in a financial time series.  They describe the 
well-established deviations from the Gaussian random walk, such as an approximate 
scaling and heavy tails of the return distributions, long-ranged volatility-volatility 
correlations (volatility clustering) and return-volatility correlations (leverage effect).  
The model is tested successfully to fit joint distributions of the 100+ years of daily 
price returns of the Dow Jones 30 Industrial Average.
\end{abstract}


\end{frontmatter}

The methods developed in studying complex physical systems have been
successfully applied throughout decades to analyze financial data \cite
{bachelier,levy,mandelbrot} and they continue to attract gradual interest 
\cite{cont,vandewalle,para,ales,fama,mantegna,holyst}. The field of research
connected to modeling financial markets and development of statistically
based real-time decision systems has recently been named Econophysics. In
this paper, we construct a phenomenological model for description of the
multivariate distribution of returns in a financial time series.

The random walk model proposed by Bachelier in the year 1900 \cite{bachelier}
is equivalent to the Gaussian multivariate probability density function
(PDF) of the returns $\xi _{i}$: 
\begin{equation}
G_{n}(\xi _{i})=\frac{1}{(2\pi )^{n/2}}\exp (-\frac{1}{2}\sum_{i=1}^{n}%
\sigma _{i}^{-2}\xi _{i}^{2})\prod_{i=1}^{n}\frac{1}{\sigma _{i}}.
\label{gauss}
\end{equation}
The absence of the correlations, $\mathrm{Corr}[\xi _{i},\xi _{j}]=0,$ 
from a time frame greater than $t_{i}-t_{i-1}=$ $20$ Min \cite{para}
has been widely documented and is often cited as a support for the
efficient market hypothesis \cite{fama}. The multivariate PDFs constructed in this work are
extensions of the Gaussian PDFs, aimed to model the well-established
deviations in the behavior of financial time series from the Gaussian random
walk.

The L\'{e}vy stable truncated univariate PDFs \cite{levy,mantegna} are known
to provide, for a financial time series, (i) an approximate scaling
invariance of the univariate PDFs with a slow convergence to the Gaussian
behavior and (ii) the existence of heavy tails. We propose the multivariate
Student PDFs, 
\begin{equation}
S_{n}^{\alpha }(\xi _{i})=\frac{\Gamma (\frac{\alpha +n}{2})}{(\alpha \pi
)^{n/2}\Gamma (\frac{\alpha }{2})}(1+\frac{1}{\alpha }\sum_{i=1}^{n}\omega
_{i}^{-2}\xi _{i}^{2})^{-\frac{\alpha +n}{2}}\prod_{i=1}^{n}\frac{1}{\omega
_{i}},  \label{St}
\end{equation}
for modeling the empirical PDFs with returns $\xi _{i}$. The marginal PDF
(\ref{St}) is again PDF (\ref{St}). If we integrate out all of the $\xi _{i}$
except for one, we get (\ref{St}) with $n=1$. The tails of the distributions
behave empirically like \cite{para} $\sim d\xi /\xi ^{4},$ and so $\alpha
\sim 3$.

For the Student PDF (\ref{St}), we have $\mathrm{E}[\xi _{i}]=0$ and $%
\mathrm{Corr}[\left| \xi _{i}\right| ,\left| \xi _{k}\right| ]=2/(2+\pi
)=\allowbreak \,0.39.$ The correlation of the returns vanishes as in the
Gaussian random walk. The square of the volatility equals $\sigma _{i}^{2}=%
\mathrm{Var}[\xi _{i}]=\omega _{i}^{2} \alpha /(\alpha -2).$ 

The cumulative returns, $\xi =\sum_{i=1}^{n}\xi _{i},$ are described
by 
\begin{equation}
dW(\xi )=\int d\xi \delta (\xi -\sum_{i=1}^{n}\xi _{i})S_{n}^{\alpha }(\xi
_{i})\prod_{i=1}^{n}d\xi _{i}=S_{1}^{\alpha }(\frac{\xi }{\Omega })\frac{%
d\xi }{\Omega }  \label{scaling}
\end{equation}
where $\Omega ^{2}=\sum_{i=1}^{n}\omega _{i}^{2}.$ The variance of the $\xi $
increases linearly with $n,$ in agreement with the Gaussian random walk and
in the rough agreement with the empirical observations. Eq.(\ref{scaling})
represents the scaling law for financial time series.

The multivariate Student PDFs have therefore heavy tails and the exact
scaling invariance from the start.
These distributions can be modified further to describe two other
well-established stylized facts which are (iii) long ranged
volatility-volatility correlations that are also known as volatility
clustering \cite{ding} and (iv) return-volatility correlations that are also
known as leverage effect \cite{black,cox}.

The empirical facts show that there is a slow decay of the correlation
function. An extension of the PDF (\ref{St}) that has the value $\mathrm{Corr%
}[\left| \xi _{i}\right| ,\left| \xi _{j}\right| ]$ which is decaying with
time is rather straightforward. We use uncorrelated multivariate
distributions for different groups of the returns. The analogy with the
Ising model can be useful: The groups $\xi _{i}$ with the same multivariate
Student PDF can be treated as domains of spins aligned in the same
direction. We assign the usual probability for every such configuration 
\begin{equation}
w[\sigma _{1},...,\sigma _{n}|\beta ]=N\exp (-\beta \sum_{i=1}^{n-1}\sigma
_{i}\sigma _{i+1})  \label{ising}
\end{equation}
where $\sigma _{i}=\pm 1.$ The normalization constant is given by $%
1/N=2(2\cosh (\beta ))^{n-1}.$ The correlation of the absolute values of the
returns equals $2/(2+\pi)$ provided that $\xi _{i}$ and $\xi _{k}$ belong
to the same domain. Otherwise the result is zero. The probability of getting
the $\xi _{i}$ and $\xi _{k}$ within the same domain can be found to be $%
w_{l}=e^{-\gamma l}$ where $e^{-\gamma }=e^{-\beta }/(e^{\beta }+e^{-\beta
})<1\,$and $l=i-k.$ The coefficient $\mathrm{Corr}[\left| \xi _{i}\right|
,\left| \xi _{k}\right| ]$ for the modified PDF has therefore the form 
$2/(2+\pi)w_{l}.$ The
absence of the correlations would formally correspond to $\beta =+\infty $ $%
(\gamma =+\infty )$. This is the case when the multivariate PDF is a product
of the $n$ univariate PDFs.

In order to incorporate leverage effect, we consider the $\omega$-function 
depending on the signs $\epsilon _{i-p}=\mathrm{sign}%
(\xi _{i-p})$ of the lagged returns ($p=1,2,$ $...$)$.$ The values $%
\epsilon _{i}=\pm 1$ are assumed to be independent variables which take the
two values $\pm 1$ with the equal probabilities. The $\epsilon _{i}$ dependence is
modeled by $\omega _{i}=C\left( 1-\rho \frac{1-\nu }{\nu }\sum_{p=1}^{\infty }\nu
^{p}\epsilon _{i-p}\right) ^{2}$ where $C$ is a normalization factor. 
Negative recent returns $\epsilon _{i-p}=-1$ ($%
p=1,2,$ $...$) increase the volatility, so $\rho >0.$ The value of $\rho $ is connected to the
overall strength of leverage effect.

Taking volatility clustering and leverage effect into account, we obtain for $l=i-k>0$: 
\begin{eqnarray}
\mathrm{Corr}[\left| \xi _{i}\right| ,\left| \xi _{k}\right| ] &=&\frac{2}{%
\pi }\frac{(1-\frac{2}{\pi })g_{l}w_{l}+\frac{2}{\pi }(g_{l}-1)}{g_{0}-\frac{%
4}{\pi ^{2}}},  \label{clus} \\
\mathrm{Corr}[\left| \xi _{i}\right| ,\xi _{k}] &=&\frac{2}{\pi }\frac{(1-%
\frac{2}{\pi })w_{l}+\frac{2}{\pi }}{\sqrt{(g_{0}-\frac{4}{\pi ^{2}})g_{0}}}%
h_{l}  \label{leve}
\end{eqnarray}
where $g_{l}=\mathrm{E}[\omega _{i}\omega _{k}]$ and $h_{l}=\mathrm{E}[\omega
_{i}\epsilon _{k}\omega _{k}]$ can be calculated analytically. 

The empirical correlation functions for the 100+ years of daily 
price returns of the Dow Jones 30 Industrial Average are
fitted using a superposition $w_{l}=c_{1}e^{-\gamma _{1}l}+c_{2}e^{-\gamma
_{2}l}\ $for$\;l>0$ with $c_{1}=0.18,$ $c_{2}=0.08,\,\gamma _{1}=1/1200,$
and $\gamma _{2}=1/233$ (else $c_{0}=0.74\,$and $\gamma _{0}=+\infty ,\,$so
that $\sum_{m=0}^{2}c_{m}=1$)$.$ We have also used $\rho =1\,$and $\nu =\exp (-1/16).$ The
results are in a good agreement with the data. The quantitative comparison can be 
found in Ref. \cite{mik}. The value of $c_{1}+c_{2}=0.26$ indicates
that $26\%$ of the empirical PDF consists of products of the multivariate Student PDFs. 
The value $\mathrm{Corr}[\xi
_{i},|\xi _{k}|]$ vanishes at $i>k$ in agreement with the observations,
since the $\omega _{i}$ depend on lagged returns only.

The complete multivariate PDF of the returns is given by 
\begin{eqnarray}
S_{n}^{\alpha }(\xi _{n},...,\xi _{1})_{C}
&=&\sum_{m=0}^{2}c_{m}\sum_{s=0}^{n-1}\frac{\exp (-\beta _{m}(n-2s-1))}{%
(2\cosh (\beta _{m}))^{n-1}}  \nonumber \\
&&\times \sum_{n_{s},...,n_{0}}\prod_{f=1}^{s+1}S_{n_{f}-n_{f-1}}^{\alpha
}(\xi _{n_{f}-1},...,\xi _{n_{f-1}})  \label{final}
\end{eqnarray}
where $n_{s+1}=n+1,$ $n\geq n_{s}>...>n_{1}\geq 2,$ and $n_{0}=1.$ 
The marginal probability of the PDF (\ref{final})\ is a PDF\ (\ref{final}) again.

The authors wish to thank the Dow Jones Global Indexes for providing 
historical prices of the Dow Jones Averages. M.I.K. is grateful to
Metronome-Ricerca sui Mercati Finanziari for kind hospitality.

% \appendix

% \section{}
% \label{}


\begin{thebibliography}{00}


% notes:

% \bibitem{label2}
% If there is a note, it should come last:
% \bibitem{label3} \note
% \end{subbibitems}

\bibitem{bachelier}  L.\ Bachelier, \textit{Ann. Sci. Ercole Norm. Suppl.} 
\textbf{3}, 21 (1900).

\bibitem{levy}  P. Levy, \textit{Th\'{e}orie de l'Addition des Variables
Al\'{e}atoires} (Gauthier-Villars, Paris, 1937).

\bibitem{mandelbrot}  B. B. Mandelbrot,\textit{\ J. Business} \textbf{36},
294 (1963).

\bibitem{cont}R. Cont and D. Sornette, \textit{J. Phys. I France} \textbf{7}, 431 (1997).

\bibitem{vandewalle} N. Vandewalle and M. Ausloos, \textit{Physica A}  \textbf{246} 454 (1997).

\bibitem{para}  P. Gopikrishnan, V. Plerou, L. A. Nunes Amaral, M. Meyer,
and H. E. Stanley, \textit{Phys. Rev.} \textbf{E60}, 5305 (1999).

\bibitem{ales}  E. Alessio, A. Carbone, G. Castelli, and V. Frappietro, 
\textit{The European Physical J. }\textbf{B27, }197 (2002).

\bibitem{fama}  E. F. Fama, \textit{Journal of Finance} \textbf{25, }383
(1970).

\bibitem{mantegna}  R. N. Mantegna and H. E. Stanley, \textit{Nature} 
\textbf{376,} 46 (1995).

\bibitem{holyst} K. Urbanowicz and J. A. Holyst, \textit{Phys. Rev.} \textbf{E67}, 046218  (2003). 

\bibitem{ding}  Z. Ding, C. W. J. Granger and R. F. Engle, \textit{J.
Empirical Finance} \textbf{1}, 83 (1993).

\bibitem{black}  F. Black, \textit{Proceedings of the 1976 Americal
Statistical Association, Business and Economical Statistics Section}, p. 177
(1976).

\bibitem{cox}  J. C. Cox and S. A. Ross, \textit{J. Fin. Eco.} \textbf{3},
145 (1976).

\bibitem{mik}  E. Alessio, V. Frappietro, M. I. Krivoruchenko, and L. J. Streckert, \\
http://arXiv.org/abs/cond-mat/0310300.

\end{thebibliography}
\end{document}